# Towards Air Stability of Ultra-Thin GaSe Devices: Avoiding Environmental and Laser-Induced Degradation by Encapsulation

Qinghua Zhao[1,2,3], Riccardo Frisenda[4,*], Patricia Gant[3], David Perez de Lara[4], Carmen Munuera[3], Mar Garcia-Hernandez[3], Yue Niu[4,5], Tao Wang[1,2,*], Wanqi Jie[1,2], Andres Castellanos-Gomez[3,*]

[1] State Key Laboratory of Solidification Processing, Northwestern Polytechnical University, Xi'an, 710072, P. R. China

[2] Key Laboratory of Radiation Detection Materials and Devices, Ministry of Industry and Information Technology, Xi'an, 710072, P. R. China

[3] Materials Science Factory. Instituto de Ciencia de Materiales de Madrid (ICMM-CSIC), Madrid, E-28049, Spain.

[4] Instituto Madrileño de Estudios Avanzados en Nanociencia (IMDEA-nanociencia), Campus de Cantoblanco, E-28049 Madrid, Spain.

[5] National Key Laboratory of Science and Technology on Advanced Composites in Special Environments, Harbin Institute of Technology, Harbin, China

*E-mail: riccardo.frisenda@csic.es , taowang@nwpu.edu.cn , andres.castellanos@csic.es

**ABSTRACT:** Gallium selenide (GaSe) is a novel two-dimensional material, which belongs to the layered III-VIA semiconductors family and attracted interest recently as it displays single-photon emitters at room temperature and strong optical non-linearity. Nonetheless, few-layer GaSe is not stable under ambient conditions and it tends to degrade over time. Here we combine atomic force microscopy, Raman spectroscopy and optoelectronic measurements in photodetectors based on thin GaSe to study its long-term stability. We found that the GaSe flakes exposed to air tend to decompose forming firstly amorphous selenium and $Ga_2Se_3$ and subsequently $Ga_2O_3$. While the first stage is accompanied by an increase in photocurrent, in the second stage we observe a decrease in photocurrent which leads to the final failure of GaSe photodetectors. Additionally, we found that the encapsulation of the GaSe photodetectors with hexagonal boron nitride (h-BN) can protect the GaSe from degradation and can help to achieve long-term stability of the devices.

**KEYWORDS:** two-dimensional material; gallium selenide (GaSe); degradation; Raman spectroscopy; optoelectronics.

**Introduction**

The isolation of two-dimensional (2D) materials has sparked a revolution in the material science and nano-science communities thanks to the ever growing amount of 2D materials that form a comprehensive catalogue of materials with very different electronic properties, ranging from superconductors to wide bandgap insulators.[1] Within this large family, transition metal dichalcogenides (TMDCs),[2] and in particular those based on Mo and W, have attracted large interest because of their bandgap within the visible range of the electromagnetic spectrum that allows for the fabrication of outstanding optoelectronic devices.[3] In addition, Mo- and W-based TMDCs 2D materials also demonstrated their potentials as solid-state-single photon emitter sources, which are important elements for quantum communication systems.[4] Unfortunately, these materials require cryogenic temperatures for operating.[4] Recently, gallium selenide (GaSe) emerged as an alternative to TMDCs for quantum optics experiments, as it displays single-photon emission even at room temperature.[5] GaSe is one of the members of the layered III-VIA semiconductors family, whose atoms are arranged in hexagonal lattices.[6] This material is also very promising for other applications in optics thanks to its strong nonlinear optical properties that allow for example strong second and third harmonic generation.[7] Moreover a responsivity of GaSe photodetectors up to 1200A/W has been reported, which is very appealing for optoelectronic applications.[8] However, few-layer GaSe is less stable than Mo- and W- based TMDCs and its degradation over time has been already reported both in experiments and theoretical calculations.[9]

Degradation processes in 2D materials are common and their origin is in part due to their layered structure. In fact, although the very large surface-to-volume ratio of 2D materials has been beneficially exploited for surface chemical functionalization and physisorption of molecules or nanoparticles,[10] this large surface-to-volume ratio also makes these materials very sensitive to the environmental atmosphere and to external stimuli in general. The mostly reported example of unstable 2D material is black phosphorous (BP),[11] which is a promising p-type layered material that exhibits high hole mobility at room temperature and thickness-dependent bandgap spanning from 0.35 eV to more than 1 eV.[12] Quickly after its isolation in ultrathin form, it was found that the degradation of mechanical exfoliated few-layer BP is a layer-by-layer etching process, along with an expansion of volume due to the condensation of moisture of water in the air.[11a] Photo-oxidation has been identified as the main cause of degradation and this process, which is faster in the case of thinner flakes, can be accelerated by increasing the concentration of oxygen or the illumination power.[11b]

Moreover, the air-degradation process was also reported in few-layer GaTe and MoTe$_2$.[13] Generally, the degradation in air of these compounds is described as a complicated oxidation process,[14] which comes along with the quenching of optical properties and changes in chemical composition, induced by O$_2$ and atmospheric humidity, which can be promoted by increasing environment temperature and by illuminating with above-bandgap light sources. Surprisingly, it was also found that in CVD-grown TMDCs (MoS$_2$, WS$_2$) a slow aging process is present, which is absent in mechanically exfoliated flakes, and it was attributed to grain boundaries and defect sites in the TMDCs films.[15]

The degradation process of GaSe flakes with different thickness has been investigated by various groups using Raman spectroscopy. The general observations of these studies are that thin GaSe flakes tend to degrade over time (in a time-scale that goes from hours to days) when left in air (see Table S1 for a comparison of the results present in the literature).[9a-f] The variation in degradation time reported in these studies can be explained by differences in parameters such as the thickness of samples, the number and nature of defects, or the environmental humidity. Moreover, photo-induced degradation of GaSe can even happen during the Raman spectroscopy characterization.[9a] The long integration time and high excitation laser power used in the Raman spectroscopy experiments could decrease the degradation time to only a few seconds.[16] Among these reports, encapsulation methods, such as plasma-enhanced chemical vapor deposition (PECVD, Si$_x$N$_y$, 32nm) and atomic layer deposition (ALD, Al$_2$O$_3$, 50nm),[9d, 9f] were proposed to protect the GaSe flakes and maintain their optical properties intact. While these works report the influence of the degradation on the optical properties of the materials, an experimental study of the environmental stability of the electrical and optoelectronic properties of GaSe-based photodetectors is still lacking in literature.

In this work, we present a study of the environmental and laser-induced degradation of mechanically exfoliated GaSe flakes from a high-quality GaSe bulk crystal.[17] We combine optical microscopy, atomic force microscopy (AFM), Raman spectroscopy, and optoelectronic transport measurements on GaSe photodetector devices kept under different atmospheres (air, vacuum, O$_2$ and N$_2$) to study the influence of the degradation on the surface morphology and electronic and optical properties of few-layer GaSe. This combination of techniques allows us to get a deeper insight on the degradation process and we find that both the presence of O$_2$ and of external illumination play important roles in the degradation of GaSe, which is in agreement with recent theoretical calculations.[9g] We also demonstrate how the encapsulation of a thin GaSe flake by a top h-BN flake is an effective method to prevent the degradation of the material and to increase the life time of a device from a couple of days (for un-encapsulated devices) to months (for h-BN encapsulated). These results give new insights about the photo-oxidation of GaSe and can pave the way to the development of future devices based on this material.

**Results and Discussion**

Few-layer GaSe flakes are fabricated by mechanical exfoliation with Nitto tape (SPV 224) of a bulk GaSe crystal, grown by the Bridgman method (see Figure 1a, bottom).[17] Figure 1a shows the lattice structure of GaSe where the different atomic planes bound by van der Waals forces are visible. The flakes cleaved by the tape are then transferred onto the surface of a polydimethylsiloxane (PDMS) stamp (Gelfilm by Gelpak) by gently pressing the tape against the surface of the PDMS substrate and peeling it off slowly. Figure 1b shows a transmission mode optical microscopy image of a GaSe flake on a PDMS substrate. The flakes on the PDMS can be then easily transferred onto another arbitrary substrate with micrometric precision by using a deterministic transfer setup.[18] Figure 1c shows the same flake of Figure 1b after being transferred onto a $SiO_2$/Si substrate with two pre-patterned gold electrodes. The two regions in the flake that show different colors in the transmission-mode picture of Figure 2b correspond to different thicknesses of the GaSe. The thinnest region which form the channel of the device is 14 nm thick as found from atomic microscope force (AFM) measurements, corresponding to approximately 14 layers.

After the fabrication we inspected the transferred GaSe devices by optical microscopy as a function of time to look for changes in the appearance of the flakes that would indicate their degradation. No obvious changes could be observed even after several hours of exposure to ambient conditions and the first symptoms of degradation could be observed after approximately 3 days (see Figures S2 and S5). We then studied the samples with AFM in order to probe subtle topographical changes that could not be resolved with the optical microscope. Figure 2a shows the optical microscopy image of a pristine GaSe flake onto $SiO_2$/Si whose topographical properties have been studied as a function of exposure time to ambient conditions. Figure 2b shows the topography of the region highlighted with a white dashed rectangle in panel (a). The surface of GaSe is highly uniform over a large scale with atomically flat large terraces and 1-2 nm high steps. Figure 2c shows the topography of the region highlighted in 2b with a black dashed square as a function of time. By inspecting the images, one can see that, at the initial state, the surface of the as-exfoliated GaSe presents many elongated bubbles of ~7 nm in height and an area of ~12000 $nm^2$. Such topographic features were also observed by Tonndorf et al..[5] The second AFM image, recorded after 5 hours, presents a smaller number of bubbles characterized by a circular shape and with an area of ~6000 $nm^2$ and height of ~20 nm. This topographic evolution seems to indicate that the original bubbles are probably due to adsorbates trapped between the flake and the substrate during the transfer process that diffuse at the flake/substrate interface over time to coalesce into larger bubbles. Similar behavior has been observed in graphene transferred onto h-BN after thermal annealing.[19] In the six subsequent images, recorded between 9 and 30 hours after the fabrication, the individual bubbles grow in dimensions, with the average diameter increasing from 100 nm to 500 nm and the height from 10 nm to 60 nm. From the top panel of Figure 2d, which shows the evolution of the average height of the bubble extracted from a statistical analysis of the AFM images as a function of time, the height of the bubbles steadily increases at a rate of ∼1.6 nm/h. This process can introduce strain in the atomic lattice thus modifying the optical and electronic properties of GaSe.[9g, 20] A second observation is that the number of bubbles increases in time as shown by the bottom panel of Figure 2d. The initial formation and growth of an individual bubble is shown in Figure 2e by the cross section of the AFM images taken in the region highlighted by the arrows in panel (c). We also analyzed the evolution of the AFM topography of a GaSe flake in $N_2$ atmosphere (see Supporting Information, Figure S3). We found that after 125 hours of $N_2$ exposure the surface of the flake shows a topography very similar to the one observed in the initial stages of the GaSe flakes in air (multiple elongated bubbles with small height) and no signs of high circular bubbles were found. After the 125 hours in $N_2$ atmosphere we exposed the flake to air and continued to monitor the topographic changes. Rapidly after the exposure to air the low elongated bubbles start to coalesce into high circular bubbles similarly to what is shown in Figure 2. An abrupt change in the surface roughness and in the flake height can be observed right after the exposure to air of the flake (Figure S3). We then attribute the nucleation of the high circular bubbles to the reaction with $O_2$ or $H_2O$ present in the air.

To get an insight about a possible chemical and/or structural transformation of the exfoliated GaSe associated with the observed topographical changes, Raman spectroscopy measurements were performed over a period of approximately 200 hundreds hours. Figure 3a shows a sequence of Raman spectra acquired on a 14 nm thick GaSe flake kept in air as a function of time (the flake used is the same one shown in Figure 1). The spectrum acquired from the freshly cleaved GaSe shows three prominent peaks centered at 134 cm$^{-1}$, 214 cm$^{-1}$ and 308 cm$^{-1}$. These peaks can be attributed to the $A_1'$ (1), E' (TO), and $A_1'$ (2) vibrational modes of crystalline GaSe, respectively (see Figure 3b).[21] Upon atmospheric exposure the peak at 214 cm$^{-1}$ (E' (TO)) gets less and less defined because of the development of three broad peaks related to the appearance of $Ga_2Se_3$, $Ga_2O_3$, and $α$-Se (amorphous selenium) chemical species in the sample that contribute with Raman modes centered at 155cm$^{-1}$, 202cm$^{-1}$ and 236cm$^{-1}$,[22] respectively. Figure 3c shows how the spectrum of the sample exposed to environmental conditions can be fitted to identify the presence of other species different from pristine GaSe. The evolution of these six peaks can be monitored in time during the degradation of GaSe.

In Figure 3d, we display the total area of the three peaks corresponding to pristine GaSe (134cm$^{-1}$, 214cm$^{-1}$ and 308cm$^{-1}$, labelled as 'GaSe' in the figure) and the peaks corresponding to other species that appear upon air exposure (155cm$^{-1}$, 202cm$^{-1}$, and 236cm$^{-1}$, labelled as 'aged' in the figure). The decrease of the 'GaSe' signal and the increase of the 'aged' one indicate that the GaSe is being converted into $Ga_2Se_3$, $\alpha$-Se, and $Ga_2O_3$. Interestingly, we found that these 'aged' species do not appear all at the same time. Figure 3e displays the time evolution of the area of the peaks associated to 155cm$^{-1}$, 202cm$^{-1}$, and 236cm$^{-1}$ demonstrating that while $\alpha$-Se and $Ga_2Se_3$ appear almost right after the exfoliation, and $Ga_2O_3$ starts showing up and growing after that the sample was kept for approximately 40 hours in air.

Raman measurements similar to the ones discussed above have the risk of damaging the GaSe flakes, thus introducing artifacts in the interpretation of the environmental degradation data. The left panel of Figure 4a shows a sequence of Raman spectra acquired upon increasing integration time while keeping the incident power density 0.11 mW/µm$^2$, on a 14 nm thick GaSe flake deposited on SiO$_2$/Si substrate. For short integration time the spectra only show the features corresponding to pristine GaSe, but for integration of 20 s or more the spectra start to develop new broad features and after the measurement the surface of the flake seems to have been damaged as can be seen in the microscope picture shown in Figure 4b. This fast degradation can help to explain previous results that reported degradation times on the order of few minutes when using a high excitation power laser and/or a long integration time.[9a, 9c] We found that by transferring the flake onto a gold surface this laser-induced degradation can be reduced. The central panel of Figure 4a shows how for the GaSe flake on gold integration time of up to 30 s does not show up any laser–induced degradation. This could be related to a more efficient heat management because of the higher thermal conductivity of gold. However, for longer integration times the flakes suffer a similar laser-induced ablation (see the optical image in Figure 4b) and the spectra develop again extra features not expected for pristine GaSe. Moreover, apart from the differences in laser-induced degradation time on Au and SiO$_2$ we also observe that the broad peak due to amorphous selenium (for GaSe/Au centered at 236 cm$^{-1}$) shifts at higher energies in the case of GaSe/SiO$_2$ (256 cm$^{-1}$), in the spectra recorded with an integration time larger than 60s. Such phenomenon can be explained by photo-crystallization of amorphous selenium.[9a, 9b, 23] Interestingly, we have found that a partial encapsulation with h-BN is an effective method to completely prevent this laser-induced damage as evidenced by the Raman spectra shown in the right panel of Figure 4a. Figure 4c illustrates how a GaSe deposited on Au and capped with a h-BN flake does not present any laser-induced damage even for the longer integration times used in our experiment. This result suggests that the laser-induced degradation requires the presence of air, indicative of a photo-oxidative process. A further comparison is presented in Figure S4, under the same laser excitation power and with the increasing integration time, the photo-induced damage level on the GaSe structure on SiO$_2$ and Au/SiO$_2$ substrate in air is much higher than in the h-BN/GaSe/Au geometry.

In the following, we study the environmental exposure effect on the optoelectronic properties of GaSe devices. We fabricate GaSe photodetectors by directly transferring GaSe flakes bridging two gold electrodes pre-patterned on a SiO$_2$/Si substrate as shown in Figure 1c. Figure 5a shows the current-voltage (*I-V*) characteristics of one GaSe device. In the dark state, the device is highly resistive and its dark current is within the noise level of our setup showing a resistance of 420 GΩ. Upon global illumination with light at 405 nm of wavelength and a power density of 1.02 W/cm$^2$ the device builds up a sizeable photocurrent with the *I-V*s being linear and symmetric in bias voltage. As can be seen in the plot after 72 hours in ambient conditions the device generates more photocurrent than the pristine one while after 144 hours the photocurrent largely decreased. The resistance under illumination of the device is 1.5 GΩ, 1.1 GΩ and 20 GΩ in the pristine, after 72 hours and 144 hours, respectively. The photoresponse of the device to modulated illumination (*I-t* curves), recorded with a bias of 8 V, is plotted in Figure 5b.

To further investigate the role of the environmental induced degradation in the optoelectronic performance of the devices we follow the long-term evolution of the photocurrent of devices kept in different environmental conditions. Figure 5c shows the time evolution of the photocurrent ($I_t$) normalized to the photocurrent of the pristine device recorded right after fabrication ($I_0$) for four GaSe photodetectors kept in different environmental conditions. Interestingly, upon air exposure the photocurrent increases steadily until almost triplicating its initial value at 39.5 hours. Importantly this change in photocurrent occurs before the appearance of Ga$_2$O$_3$ in the device revealed by Raman spectroscopy (the device tested in air exposure is the same one studied in Figures 1 and 3). The initial increase of photocurrent is also observed in a device measured in pure O$_2$ atmosphere, indicating that O$_2$ plays a critical role on the initial increase of the photocurrent. Indeed, according to our previous electronic probe microanalysis (EPMA) characterization,[17] the atom ratio of Se and Ga is slightly smaller than 1 (Se:Ga = 49.5:50.5) in GaSe bulks, which indicates that the dominant defects in the pristine GaSe flakes are given by Se vacancies that may act as effective sites for O$_2$ adsorption.[9g, 24]

After the initial increase, the photocurrent saturates between approximately 40 and 70 hours and then it decreases. At the same time the Raman spectra show a reduction of the amount of GaSe present in the device and the appearance of Ga$_2$O$_3$ among the aged components. Being a wide bandgap insulator (the bandgap is approximately 5 eV in bulk), Ga$_2$O$_3$ is not expected to yield photoresponse at 405 nm illumination and thus the conversion of GaSe to $Ga_2O_3$ is expected to reduce the total photocurrent generated in the device.[25] The continuous oxidation process leads to the further increasing of $Ga_2O_3$, which results in the drop of photocurrent and the final device failure. We also studied the photocurrent evolution of a device kept in vacuum (5×10$^{-4}$ mbar) and a device that has been capped by transferring a h-BN flake right after fabrication. In both cases the initial increase of photocurrent is not observed which agrees with the importance of $O_2$ to observe this photocurrent increase and with the AFM measurements of Figure 2. Interestingly, while the device measured in vacuum shows a small (but noticeable) decrease of photocurrent over time, the h-BN encapsulated device displays a constant photocurrent. We attribute the slow decrease of the photocurrent of the device in vacuum to its exposure to gases coming from the outgassing at the surrounding surfaces contrarily to the device encapsulated with h-BN that has a physical protective barrier. Remarkably, the h-BN encapsulated GaSe photodetector shows a performance comparable to the initial device even after 40 days after its fabrication (see Figure S6 and Figure S7 in the Supporting Information). This solution is quite convenient if we compare it with fully encapsulated devices, sandwiched between two h-BN flakes, that requires more complex fabrication process to contact electrically the sandwiched flakes.

To gain further information on the degradation of GaSe we studied the power dependency of the photocurrent as a function of time in a second GaSe photodetector kept in air. Figure 5d shows the results of such power dependency measurements performed on the pristine device. As can be seen from the graph, the photocurrent generated in the device grows when increasing the incident optical power. The data plotted in a log-log scale appears linear, which is a sign of a power law dependency. We fit the data to the equation $I_{ph} = A \cdot P^\alpha$, where $\alpha$ is a dimensionless fitting parameter and $A$ is a quantity related to the responsivity. The power exponent $\alpha$ contains information about the traps present in the system and is expected to be equal to 1 in the case of a traps-free photodetector and becomes smaller in the presence of traps.[26] Figure 5e shows the evolution of the exponent $\alpha$ (left axis) together with the photocurrent recorded in the device at a fixed optical power (right axis). As can be seen from the plot, the photocurrent increases in the first 100 hours and then decreases, which is in agreement with the device discussed previously. On the other hand, $\alpha$ displays a monotonous decreasing behavior, passing from 0.84 to 0.42 in approximately 200 hours. The observed behavior of $\alpha$ can be explained by an increase in the number of trap states between the valence and conduction bands as a function of time.[26b] Finally, we studied the wavelength-dependency of the photocurrent and its evolution over time. Figure 5f shows three photocurrent spectra recorded in the pristine device and after 89 hours and 138 hours of exposition to air. At short wavelengths the photocurrent can reach hundreds of pA and it rapidly decreases for longer wavelength. Eventually it goes to zero for wavelengths larger than 800 nm. Despite the change in magnitude of the photocurrent, the three spectra feature a similar shape suggesting that the photocurrent generation is dominated by the absorption of GaSe.

The results that we discussed until now indicate that degradation process of thin GaSe has very important consequences on the properties of optoelectronic devices based on this material. The Raman measurements discussed in Figures 3 and 4 indicate that GaSe degrades into various components, mainly α-selenium, $Ga_2Se_3$ and $Ga_2O_3$. Since the photocurrent spectrum of the device is dominated at all the stages by the GaSe photon absorption (as can be seen in Figure 5f), we propose that the non-monotonous evolution of the photocurrent in devices kept in air is due to a competition between various mechanisms that can influence the photocurrent: (1) the decrease of the GaSe present in the device and the increase of $Ga_2O_3$ can lead to a decrease of the total photocurrent, (2) adsorbed oxygen and selenium vacancies can increase or decrease the photocurrent due to photogating effects, (3) the appearance of $Ga_2Se_3$ and α-Se can increase the photoresponse due to built-in electric-fields and/or the introduction of strain the GaSe lattice.

Recently, Shi *et al.* reported calculations showing that photo-excited electrons in GaSe can be effectively transferred to $O_2$ molecules adsorbed on the surface and create $O_2^-$ anions.[9g] The anions can then react with GaSe and replace Ga-Se bonds by Ga-O bonds leading to the formation of the oxidation products $Ga_2O_3$ and elemental Se. This process is consistent with the Raman measurements of GaSe of Figure 3 and can explain the evolution of the photocurrent observed in Figure 5c and 5e. The trapping of electrons by $O_2$ molecules can induce an increase of the photocurrent thanks to the photogating effect (2), as observed in the first 100 hours of exposition of GaSe to air. The influence of photogating on the photoresponse of the device is also confirmed by the decrease of the power exponent *α* as a function of time. Part of this initial increase in photocurrent may also be caused by the introduction of strain (3), as suggested by the AFM measurements in Figure 2. Finally, the decrease of the photocurrent after approximately 100 hours in air can be explained by a reduction of the amount of GaSe in favor of $Ga_2O_3$ because of the oxidation process (1), which has a lower photon absorption and larger electrical resistivity under illumination of 405nm light source.[25]

**Conclusions**

In summary, we presented a study of the environmental and laser-induced degradation of ultra-thin GaSe flakes. We used complementary characterization techniques to get information about the degradation process. We combine optical microscopy, AFM and Raman spectroscopy with optoelectronic measurements on GaSe photodetectors exposed to different environmental atmosphere. We found that the environmental degradation takes place in two main stages. At first, the exposure to air induces degradation of the pristine GaSe creating α-Se and $Ga_2Se_3$, accompanied by a photocurrent increase. In the second stage, $Ga_2O_3$ appears and its concentration increases in the samples resulting in a drop of the photocurrent leading to the final failure of the GaSe photodetectors. We also found that capping the exfoliated GaSe flakes with a top h-BN flake is an effective way to passivate it, preventing its environmental degradation as well laser-induced damage.

**Materials and Methods**

**Sample fabrication.** Few-layer GaSe flakes were mechanically exfoliated from an high-quality bulk single crystals[17] by using scotch tape and Nitto tape (Nitto Denko® SPV 224) firstly and then onto a polydimethylsiloxane substrate (PDMS). After optical inspection, the selected flake was transferred from the PDMS to the pre-patterned Au (60nm)/$SiO_2$ (280nm)/Si substrate (Osilla®) as the photodetector devices and Raman samples by an all-dry deterministic transfer method.[18, 27] The gap between the parallel electrodes is 10 μm.

**Raman spectroscopy.** A Bruker Senterra confocal Raman microscopy setup (Bruker Optik, Ettlingen, Germany) was used for the sample degradation analysis with a laser excitation of 0.2 mW at 532 nm focused in a 1 μm spot. The integration time used is 20 s, while 5-80 s was used for the laser-induced degradation experiment.

**Optoelectronic properties.** Few-layer GaSe photodetectors are characterized in a homebuilt air-pressure (room temperature) probe station. The electrical measurements (*I-V*) were performed with a source-meter source-measure unit (Keithley 2450). The light source is provided by a light emitting diode (LEDD1B – T-Cube LED driver, Thorlabs) with wavelength 405 nm, coupled to a multimode optical fiber at the LED source and projected onto the sample surface by a zoom lens, creating a light spot on the sample with the diameter of 200 μm. To improve the contact between the GaSe flake and Au electrodes, all the devices used for the optoelectronic measurement have been annealed in a vacuum quartz tube with the temperature 250°C for 3 hours. Both optical and Raman characterization show that there is no change of the GaSe flakes before and after annealing (see Figure S1).

**AFM measurements.** A commercial AFM system and software (WSxM) from Nanotec,[28] operating at room temperature in air or $N_2$ environment conditions, was employed for surface morphology evolution analysis. Topographic images were acquired in contact mode (Figure 2) and 'tapping' mode (Figure S3). Commercial AFM tips from Next-Tip S.L. were used.

## ACKNOWLEDGEMENTS

This project has received funding from the European Research Council (ERC) under the European Union's Horizon 2020 research and innovation programme (grant agreement n° 755655, ERC-StG 2017 project 2D-TOPSENSE). EU Graphene Flagship funding (Grant Graphene Core 2, 785219) is acknowledged. RF acknowledges support from the Netherlands Organization for Scientific Research (NWO) through the research program Rubicon with project number 680-50-1515. DPdL acknowledges support from MINECO through the program FIS2015-67367-C2-1-p. QHZ acknowledges the grant from Chinese Scholarship Council (CSC) under No. 201700290035. TW acknowledges support from the National Key R&D Program of China: No. 2016YFB0402405.

## COMPETING INTERESTS

The authors declare no competing financial interests.

## FUNDING

Netherlands Organization for Scientific Research (NWO): Rubicon 680-50-1515

EU H2020 European Research Council (ERC): ERC-StG 2017 755655

EU Graphene Flagship: Grant Graphene Core 2, 785219

MINECO: FIS2015-67367-C2-1-p

National Key R&D Program of China: No. 2016YFB0402405

## REFERENCES

[1]     a) K. S. Novoselov, A. K. Geim, S. V. Morozov, D. Jiang, Y. Zhang, S. V. Dubonos, I. V. Grigorieva, A. A. Firsov, *science* **2004**, 306, 666; b) S. Z. Butler, S. M. Hollen, L. Cao, Y. Cui, J. A. Gupta, H. R. Gutiérrez, T. F. Heinz, S. S. Hong, J. Huang, A. F. Ismach, *ACS nano* **2013**, 7, 2898; c) R. Frisenda, A. J. Molina-Mendoza, T. Mueller, A. Castellanos-Gomez, H. S. van der Zant, *Chemical Society Reviews* **2018**.
[2]     M. Chhowalla, Z. Liu, H. Zhang, *Chemical Society Reviews* **2015**, 44, 2584.
[3]     K. F. Mak, J. Shan, *Nature Photonics* **2016**, 10, 216.
[4]     a) Y.-M. He, G. Clark, J. R. Schaibley, Y. He, M.-C. Chen, Y.-J. Wei, X. Ding, Q. Zhang, W. Yao, X. Xu, *Nature nanotechnology* **2015**, 10, 497; b) Z. He, X. Wang, W. Xu, Y. Zhou, Y. Sheng, Y. Rong, J. M. Smith, J. H. Warner, *ACS nano* **2016**, 10, 5847; c) J. Kern, I. Niehues, P. Tonndorf, R. Schmidt, D. Wigger, R. Schneider, T. Stiehm, S. Michaelis de Vasconcellos, D. E. Reiter, T. Kuhn, *Advanced Materials* **2016**, 28, 7101.

[5] P. Tonndorf, S. Schwarz, J. Kern, I. Niehues, O. Del Pozo-Zamudio, A. I. Dmitriev, A. P. Bakhtinov, D. N. Borisenko, N. N. Kolesnikov, A. I. Tartakovskii, *2D Materials* **2017**, 4, 021010.

[6] N. C. Fernelius, *Progress in Crystal Growth and Characterization of materials* **1994**, 28, 275.

[7] a) X. Zhou, J. Cheng, Y. Zhou, T. Cao, H. Hong, Z. Liao, S. Wu, H. Peng, K. Liu, D. Yu, *Journal of the American Chemical Society* **2015**, 137, 7994; b) W. Jie, X. Chen, D. Li, L. Xie, Y. Y. Hui, S. P. Lau, X. Cui, J. Hao, *Angewandte Chemie International Edition* **2015**, 54, 1185; c) X.-T. Gan, C.-Y. Zhao, S.-Q. Hu, T. Wang, Y. Song, J. Li, Q.-H. Zhao, W.-Q. Jie, J.-L. Zhao, *Light: Science & Applications* **2018**, 7, 17126; d) L. Karvonen, A. Säynätjoki, S. Mehravar, R. D. Rodriguez, S. Hartmann, D. R. Zahn, S. Honkanen, R. A. Norwood, N. Peyghambarian, K. Kieu, *Scientific reports* **2015**, 5, 10334.

[8] Y. Cao, K. Cai, P. Hu, L. Zhao, T. Yan, W. Luo, X. Zhang, X. Wu, K. Wang, H. Zheng, *Scientific reports* **2015**, 5, 8130.

[9] a) A. Bergeron, J. Ibrahim, R. Leonelli, S. Francoeur, *Applied Physics Letters* **2017**, 110, 241901; b) T. E. Beechem, B. M. Kowalski, M. T. Brumbach, A. E. McDonald, C. D. Spataru, S. W. Howell, T. Ohta, J. A. Pask, N. G. Kalugin, *Applied Physics Letters* **2015**, 107, 173103; c) M. Rahaman, R. D. Rodriguez, M. Monecke, S. A. Lopez-Rivera, D. R. Zahn, *Semiconductor Science and Technology* **2017**, 32, 105004; d) D. Pozo-Zamudio, S. Schwarz, J. Klein, R. Schofield, E. Chekhovich, O. Ceylan, E. Margapoti, A. Dmitriev, G. Lashkarev, D. Borisenko, *arXiv preprint arXiv:1506.05619* **2015**; e) D. Andres-Penares, A. Cros, J. P. Martínez-Pastor, J. F. Sánchez-Royo, *Nanotechnology* **2017**, 28, 175701; f) J. Susoma, J. Lahtinen, M. Kim, J. Riikonen, H. Lipsanen, *AIP Advances* **2017**, 7, 015014; g) L. Shi, Q. Li, Y. Ouyang, J. Wang, *Nanoscale* **2018**.

[10] a) M. Chhowalla, H. S. Shin, G. Eda, L.-J. Li, K. P. Loh, H. Zhang, *Nature chemistry* **2013**, 5, 263; b) D. Boukhvalov, M. Katsnelson, *Nano letters* **2008**, 8, 4373.

[11] a) J. O. Island, G. A. Steele, H. S. van der Zant, A. Castellanos-Gomez, *2D Materials* **2015**, 2, 011002; b) A. Favron, E. Gaufrès, F. Fossard, A.-L. Phaneuf-L'Heureux, N. Y. Tang, P. L. Lévesque, A. Loiseau, R. Leonelli, S. Francoeur, R. Martel, *Nature materials* **2015**, 14, 826; c) J. D. Wood, S. A. Wells, D. Jariwala, K.-S. Chen, E. Cho, V. K. Sangwan, X. Liu, L. J. Lauhon, T. J. Marks, M. C. Hersam, *Nano letters* **2014**, 14, 6964.

[12] a) L. Li, Y. Yu, G. J. Ye, Q. Ge, X. Ou, H. Wu, D. Feng, X. H. Chen, Y. Zhang, *Nature nanotechnology* **2014**, 9, 372; b) A. Castellanos-Gomez, L. Vicarelli, E. Prada, J. O. Island, K. Narasimha-Acharya, S. I. Blanter, D. J. Groenendijk, M. Buscema, G. A. Steele, J. Alvarez, *2D Materials* **2014**, 1, 025001; c) F. Xia, H. Wang, Y. Jia, *Nature communications* **2014**, 5, 4458; d) J. Qiao, X. Kong, Z.-X. Hu, F. Yang, W. Ji, *Nature communications* **2014**, 5, 4475.

[13] a) S. Yang, H. Cai, B. Chen, C. Ko, V. O. Özçelik, D. F. Ogletree, C. E. White, Y. Shen, S. Tongay, *Nanoscale* **2017**, 9, 12288; b) B. Chen, H. Sahin, A. Suslu, L. Ding, M. I. Bertoni, F. Peeters, S. Tongay, *ACS nano* **2015**, 9, 5326.

[14] G. Wang, R. Pandey, S. P. Karna, *Wiley Interdisciplinary Reviews: Computational Molecular Science* **2017**, 7.

[15] J. Gao, B. Li, J. Tan, P. Chow, T.-M. Lu, N. Koratkar, *ACS nano* **2016**, 10, 2628.

[16] M. Currie, J. D. Caldwell, F. J. Bezares, J. Robinson, T. Anderson, H. Chun, M. Tadjer, *Applied Physics Letters* **2011**, 99, 211909.

[17] T. Wang, J. Li, Q. Zhao, Z. Yin, Y. Zhang, B. Chen, Y. Xie, W. Jie, *Materials* **2018**, 11, 186.

[18] A. Castellanos-Gomez, M. Buscema, R. Molenaar, V. Singh, L. Janssen, H. S. Van der Zant, G. A. Steele, *2D Materials* **2014**, 1, 011002.

[19] a) S. Haigh, A. Gholinia, R. Jalil, S. Romani, L. Britnell, D. Elias, K. Novoselov, L. Ponomarenko, A. Geim, R. Gorbachev, *Nature materials* **2012**, 11, 764; b) A. Kretinin, Y. Cao, J. Tu, G. Yu, R. Jalil, K. Novoselov, S. Haigh, A. Gholinia, A. Mishchenko, M. Lozada, *Nano letters* **2014**, 14, 3270.

[20] A. De Sanctis, I. Amit, S. P. Hepplestone, M. F. Craciun, S. Russo, *Nature communications* **2018**, 9.

[21] a) R. M. Hoff, J. Irwin, R. Lieth, *Canadian Journal of Physics* **1975**, 53, 1606; b) S. Jandl, J. Brebner, B. Powell, *Physical Review B* **1976**, 13, 686; c) T. Wieting, J. Verble, *Physical Review B* **1972**, 5, 1473.

[22] a) E. Finkman, J. Tauc, R. Kershaw, A. Wold, *Physical Review B* **1975**, 11, 3785; b) A. Yamada, N. Kojima, K. Takahashi, T. Okamoto, M. Konagai, *Japanese journal of applied physics* **1992**, 31, L186; c) T. Onuma, S. Fujioka, T. Yamaguchi, Y. Itoh, M. Higashiwaki, K. Sasaki, T. Masui, T. Honda, *Journal of Crystal Growth* **2014**, 401, 330; d) P. Carroll, J. Lannin, *Solid State Communications* **1981**, 40, 81; e) A. Baganich, V. Mikla, D. Semak, A. Sokolov, A. Shebanin, *physica status solidi (b)* **1991**, 166, 297.

[23] a) K. Ishida, K. Tanaka, *Physical Review B* **1997**, 56, 206; b) V. V. Poborchii, A. V. Kolobov, K. Tanaka, *Applied physics letters* **1998**, 72, 1167.

[24] L. Shi, Q. Zhou, Y. Zhao, Y. Ouyang, C. Ling, Q. Li, J. Wang, *The journal of physical chemistry letters* **2017**, 8, 4368.

[25] W. Y. Kong, G. A. Wu, K. Y. Wang, T. F. Zhang, Y. F. Zou, D. D. Wang, L. B. Luo, *Advanced Materials* **2016**, 28, 10725.

[26] a) A. Rose, *Physical Review* **1955**, 97, 322; b) S. Ghosh, A. Winchester, B. Muchharla, M. Wasala, S. Feng, A. L. Elias, M. B. M. Krishna, T. Harada, C. Chin, K. Dani, *Scientific reports* **2015**, 5, 11272.

[27] R. Frisenda, E. Navarro-Moratalla, P. Gant, D. P. De Lara, P. Jarillo-Herrero, R. V. Gorbachev, A. Castellanos-Gomez, *Chemical Society Reviews* **2018**, 47, 53.

[28] I. Horcas, R. Fernández, J. Gomez-Rodriguez, J. Colchero, J. Gómez-Herrero, A. Baro, *Review of scientific instruments* **2007**, 78, 013705.

**FIGURES**

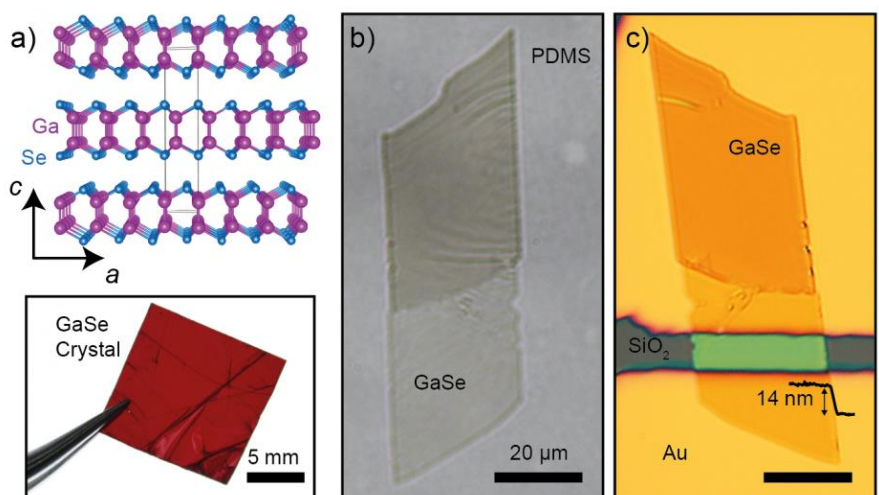

**Figure 1**: **Layered semiconducting GaSe.** a) Crystal structure of GaSe (top) and photograph of bulk GaSe (bottom). b-c) Optical pictures of few-layer GaSe exfoliated onto PDMS (b) and deterministically transferred onto pre-patterned electrodes (c).

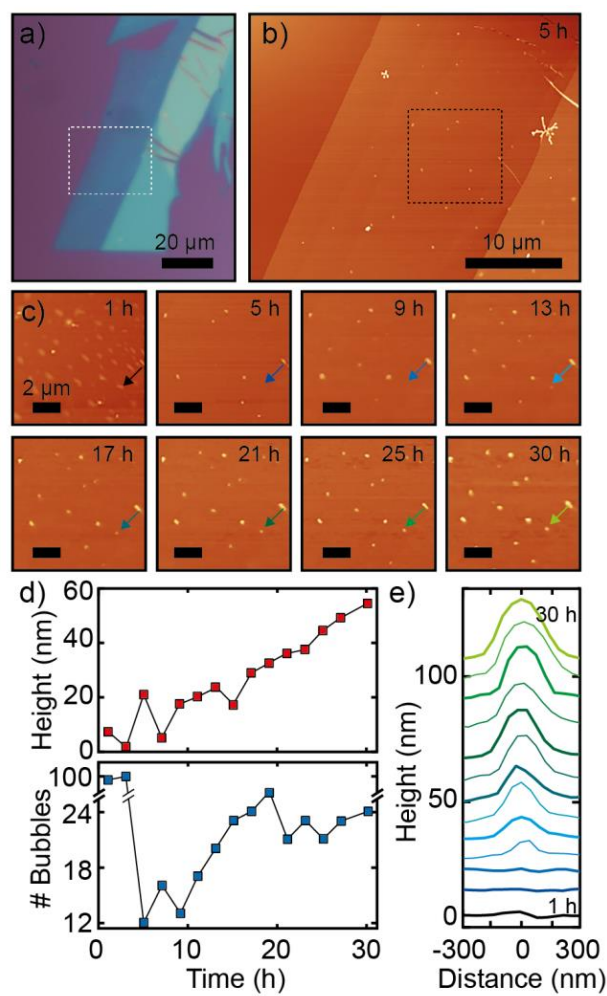

**Figure 2: Surface morphology evolution in ambient condition.** a) Optical image of a GaSe flake transferred onto the SiO$_2$/Si substrate. The area marked with white dashed square was seleted to perform the longterm (30h) AFM analysis. b) A typical surface morphology of seleted area on the GaSe flake after exposed 5 hours in air. The black dashed square area was used for following statistics analysis in panel (d). c) The surface morphology evolution of selected area on GaSe flake from 1h to 30h in air. d) The mode height and number of bubbles versus time from 1h to 30h. e) A typical evolution of cross-section shape of a bubble grown by spontanous nucleation appeared at the 9$^{th}$ hour. The selected bubble was marked by arrows with corresponding colors.

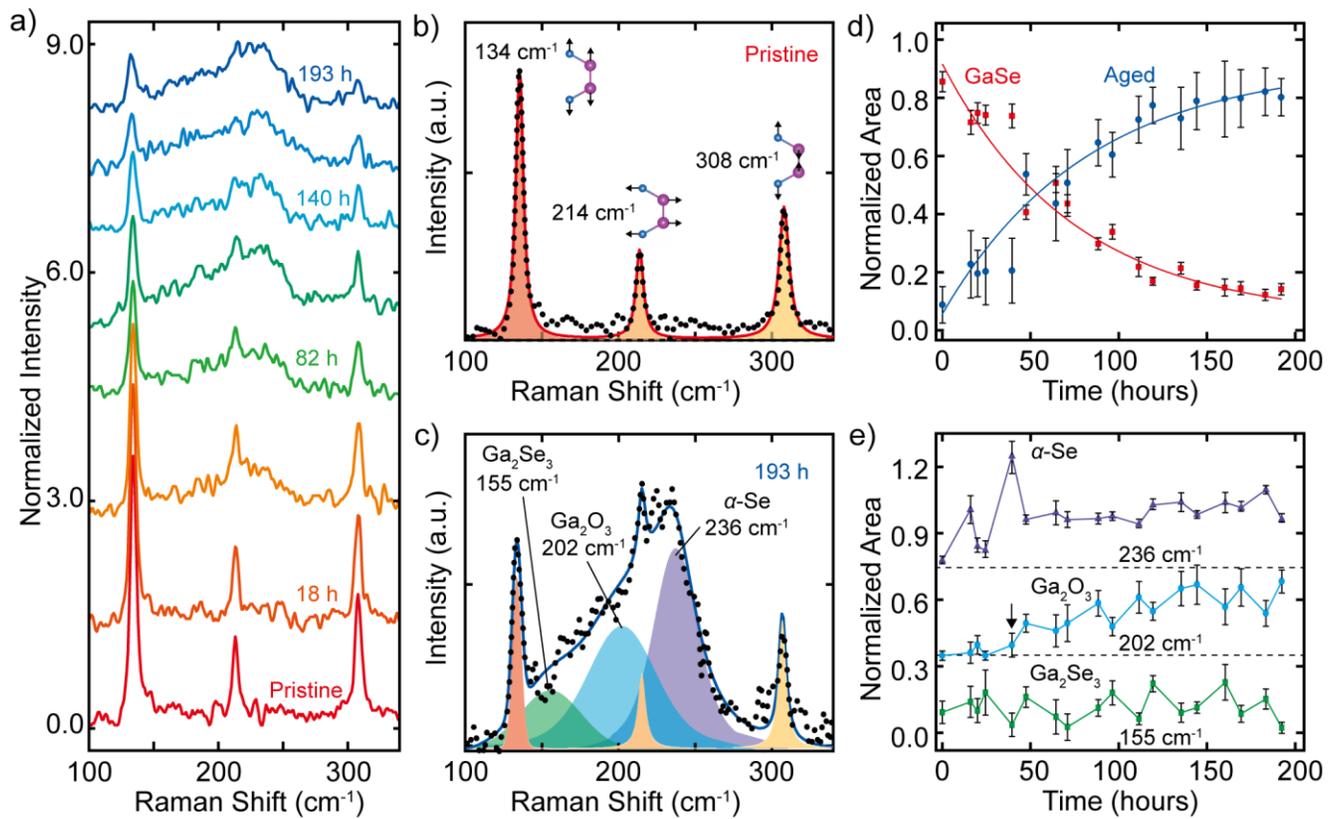

**Figure 3**: **Long-term Raman measurements of GaSe.** a) Raman spectra of 14-nm thick GaSe recorded with 532 nm laser and power density 0.11 mW/μm². The curves have been offset vertically for clarity. b-c) Raman spectra of pristine (b) and aged (c) GaSe. The thick lines represent the total fit to the data and the filled areas are the different peaks that compose each fit. d) Evolution in time of the total area of the GaSe related peaks 134 cm⁻¹, 214 cm⁻¹ and 308 cm⁻¹ (red squares) and of the aged peaks 155 cm⁻¹, 202 cm⁻¹ and 236 cm⁻¹ (blue circles). The solid lines are guidelines for the eye. e) Evolution in time of the area of the of the aged peaks 155 cm⁻¹, 202 cm⁻¹ and 236 cm⁻¹. The arrow highlights the appearance of $Ga_2O_3$ in the sample.

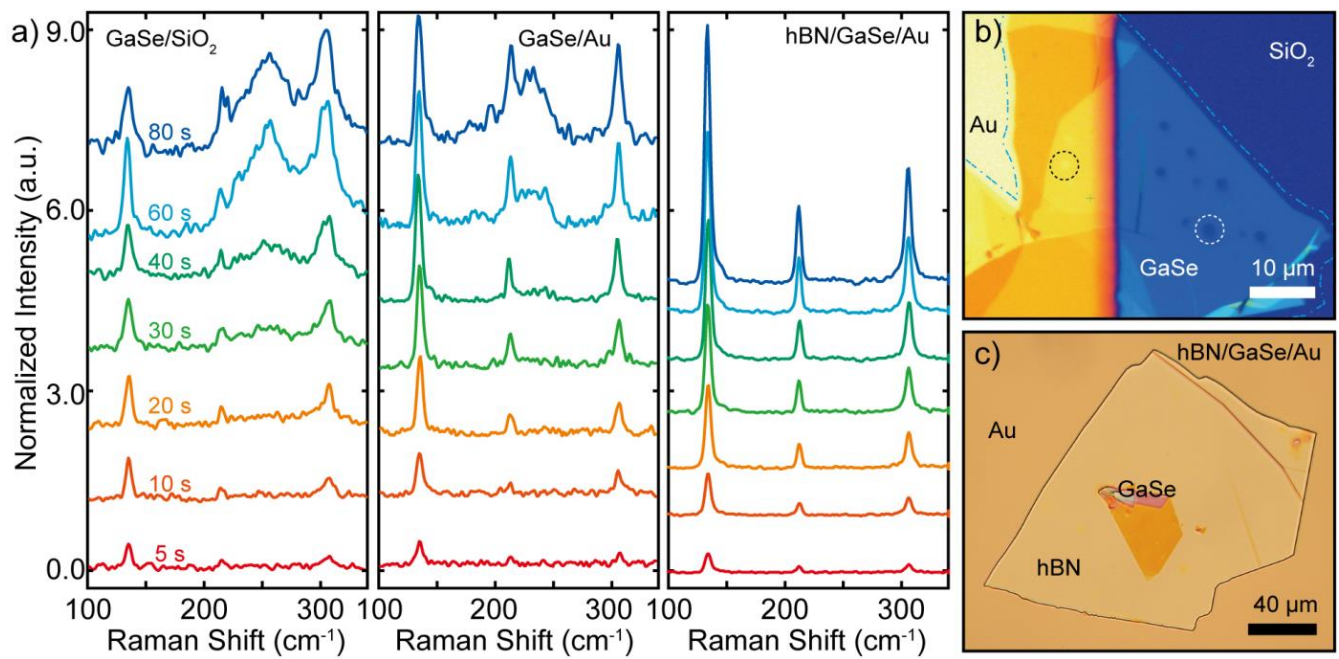

**Figure 4**: **Laser-induced degradation of GaSe.** a) Raman spectra of a GaSe flake deposited onto gold (left), SiO$_2$/Si (center) and encapsulated with a boron nitride flake. b) Microscope image of a GaSe flake deposited onto gold and SiO$_2$ captured after the laser degradation experiment. The light blue dashed line highlights the contours of the GaSe flake. The spots that are visible on the flake (two of them highlighted by the dashed lines) are induced by the laser irradiation and were absent in the pristine flake. c) Microscope image of a GaSe flake deposited onto gold and then covered by a boron nitride flake.

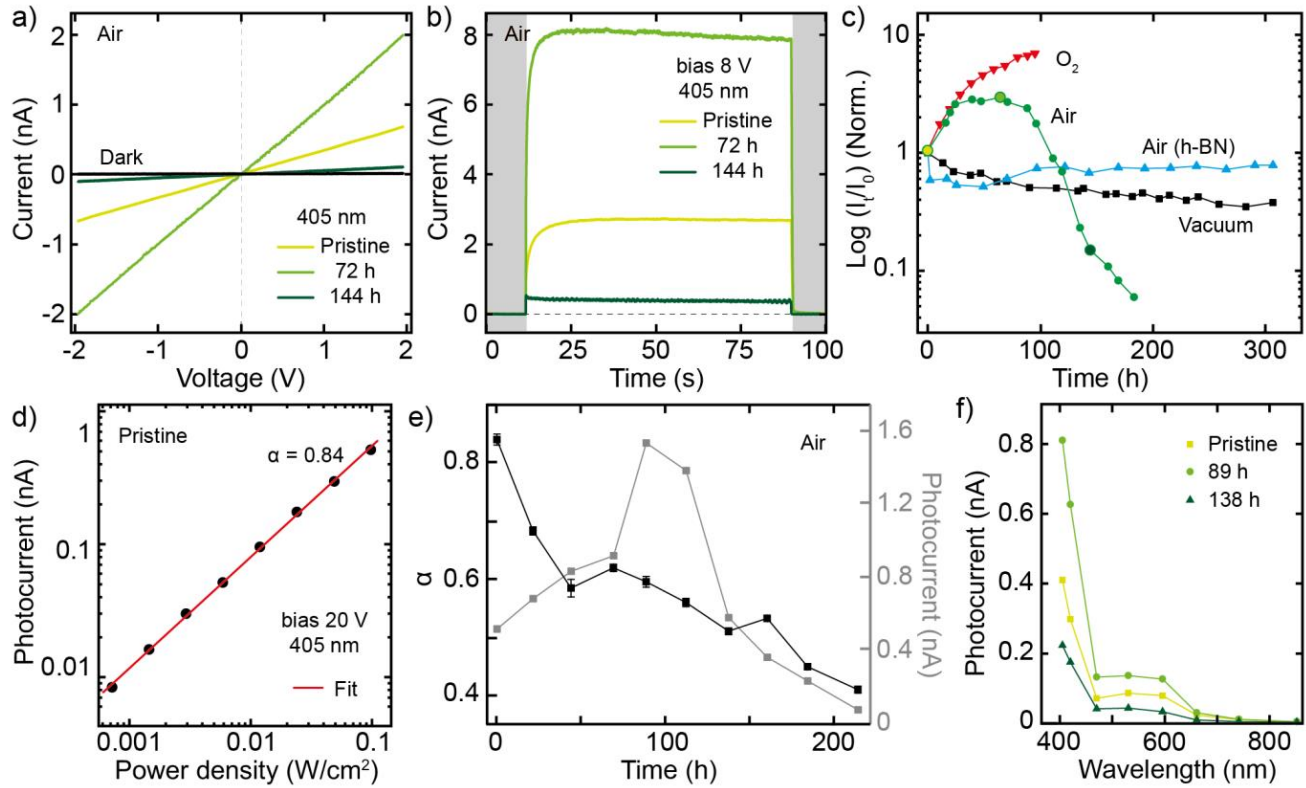

**Figure 5**: **Optoelectronic performance of GaSe photodetectors**. a) *I-V* curves of a GaSe photodetector after 0h, 72h and 144h in the air with the illumination of a 405nm light source and under dark condition. b) *I-t* curves of a GaSe photodetector kept in air after 0h, 72h and 144h. The light at 405nm is modulated with a square wave. c) Normalized photocurrent evolution versus time of GaSe photodetectors in the dry-oxygen (red), air (green, blue) and vacuum atmosphere. The blue line was obtained with a h-BN encapsulated GaSe device. d) Photocurrent of a different device as a function of incident optical power in log-log scale. The red line is a fit to a power law. e) Time evolution of the power law exponent $\alpha$ (left y-axis) and of the photocurrent (right y-axis) for the device of panel (d). f) Wavelength-resolved photocurrent for the pristine device and after being kept in air for 89 hours and 138 hours.

# Supplementary Information: Towards Air Stability of Ultra-Thin GaSe Devices: Avoiding Environmental and Laser-Induced Degradation by Encapsulation

Qinghua Zhao[1,2,3], Riccardo Frisenda[4,*], Patricia Gant[3], David Perez de Lara[4], Carmen Munuera[3], Mar Garcia-Hernandez[3], Yue Niu[4,5], Tao Wang[1,2,*], Wanqi Jie[1,2], Andres Castellanos-Gomez[3,*]

[1] State Key Laboratory of Solidification Processing, Northwestern Polytechnical University, Xi'an, 710072, P. R. China

[2] Key Laboratory of Radiation Detection Materials and Devices, Ministry of Industry and Information Technology, Xi'an, 710072, P. R. China

[3] Materials Science Factory. Instituto de Ciencia de Materiales de Madrid (ICMM-CSIC), Madrid, E-28049, Spain.

[4] Instituto Madrileño de Estudios Avanzados en Nanociencia (IMDEA-nanociencia), Campus de Cantoblanco, E-28049 Madrid, Spain.

[5] National Key Laboratory of Science and Technology on Advanced Composites in Special Environments, Harbin Institute of Technology, Harbin, China

*E-mail: riccardo.frisenda@imdea.org , taowang@nwpu.edu.cn , andres.castellanos@csic.es

| Reference | Substrate | Flake thickness | Time of complete degradation | Laser wavelength and power density used [mW/µm$^2$] |
|---|---|---|---|---|
| [9a] | SiO$_2$(100nm)/Si | 45 nm | 30 min | 532 nm, 6 (Raman) |
| [9b] | SiO$_2$(300nm)/Si | 14 nm | -- | 532 nm, 15.6 (Raman) |
| [9c] | HOPG | 3-8L | 5 h (self-limited) | 638 nm, 0.234 (Raman, 700s×3) |
| [9d] | SiO$_2$/Si | 10-25 nm | 100 h | 532 nm, 0.637 (PL) |
| [9e] | SiO$_2$(300nm)/Si | 10-40 nm | ~120 h | 532 nm, 0.127 (PL) |
| [9f] | SiO$_2$(285nm)/Si | 50 nm | > 2 weeks | 514nm, 7.643 (PL) |
| This work | Au(60nm)/SiO$_2$/Si | 14 nm | 192.5 h | 532 nm, 0.11 (Raman, 20s×2) |

**Table S1: Summary of various experiments concerning the degradation of thin GaSe.**

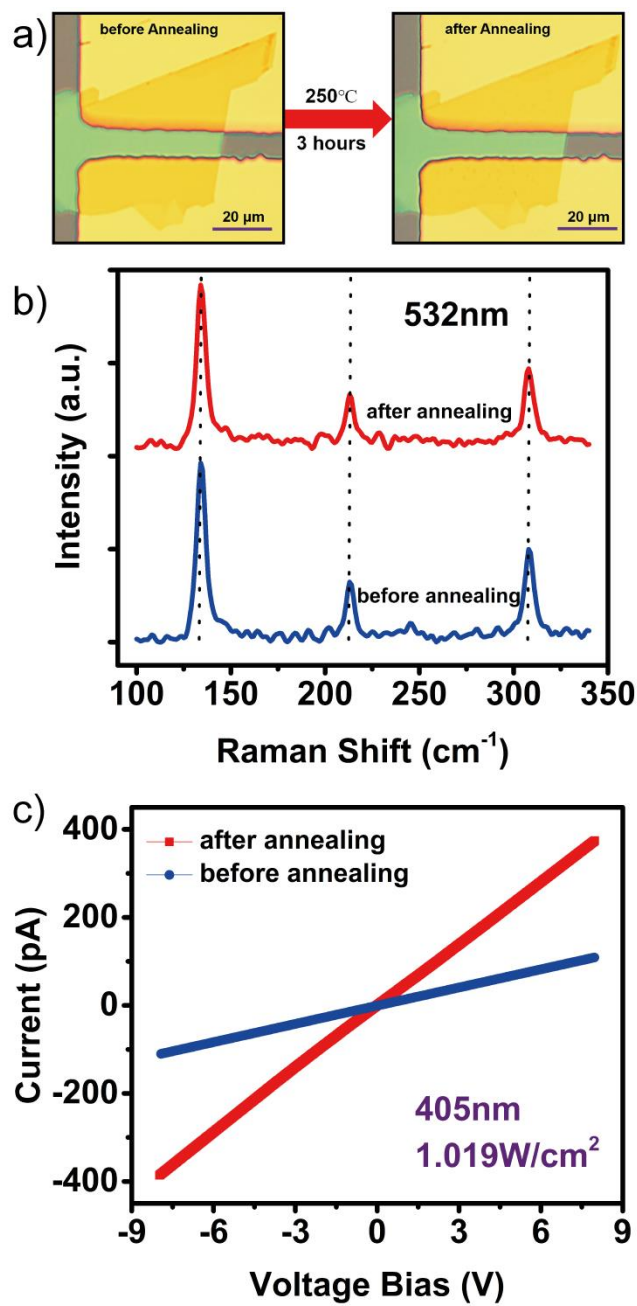

**Figure S1: Vacuum annealing of GaSe photodetectors.** Surface morphology (a), Raman spectrum (b) and *I-V* curve (c) of the device before and after annealing.

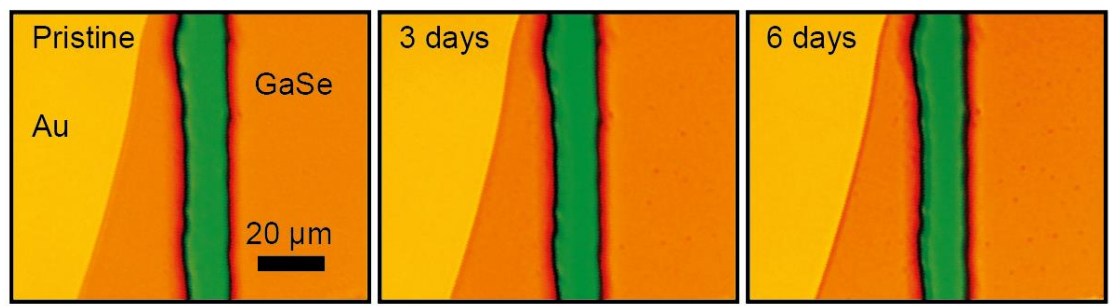

**Figure S2**: **Surface morphology evolution of GaSe in air.** Sequence of optical images of a GaSe flake kept in air. The contrast has been artificially enhanced.

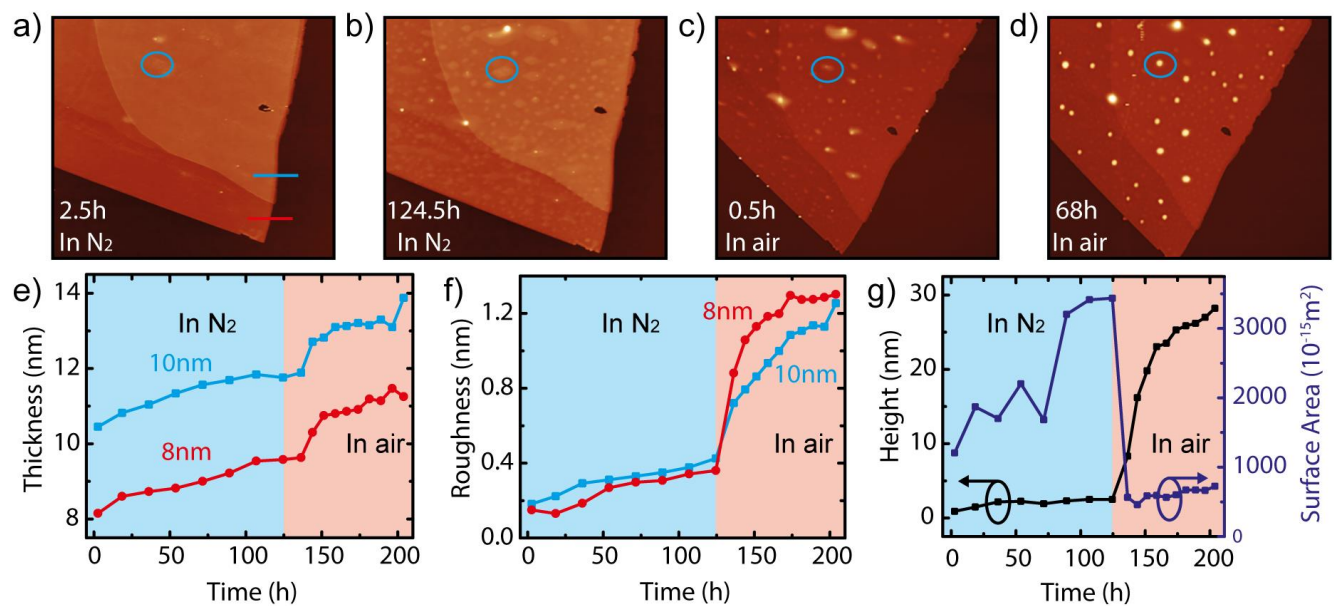

**Figure S3: Surface morphology evolution in time of GaSe flake in $N_2$ and then in air.** a-d) The surface morphology of a GaSe flake on $SiO_2$/Si substrate kept in $N_2$ atmosphere for 2.5h (a), 124.5h (b) and then in air for 0.5 h (c) and 68 h (d). The blue circles highlight a typical bubble. e-f) Thickness (e) and roughness (f) variation in time of the thinner (8nm) and the thicker (10nm) part of the GaSe flake as a function of time. g) The evolution of height and surface area of the bubbles as a function of time.

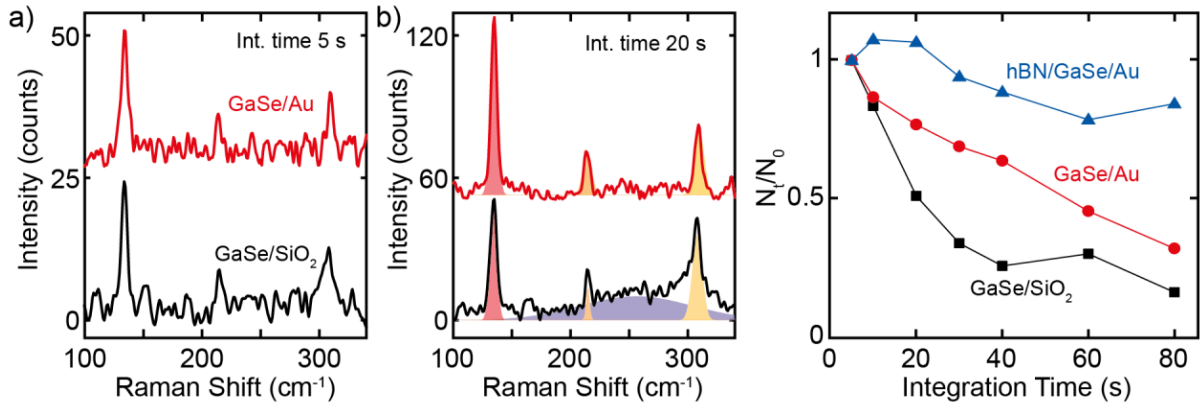

**Figure S4**: **Laser-induced degradation speed on different substrates of GaSe.** Raman spectra of a GaSe flake on Au and SiO$_2$/Si with the integration time 5s (a) and 20s (b), respectively. c) Time-dependent normalized population of emitted photons per unit time of Raman signal from the GaSe flake deposited onto gold, SiO$_2$/Si and encapsulated with a boron nitride flake. N$_0$ and N$_t$ is the population of emitted photon per unit time with the integration time 5s and the increasing time (10s, 20s, 40s, 60s, 80s).

The minor difference between the Raman intensities of the GaSe flake on Au and on SiO$_2$ with integration time 5s (a) indicates that there is no enhancement of the GaSe Raman signal on Au. Hence, with integration time 20s, the observable difference between the Raman intensity can be attributed to the faster degradation on SiO$_2$ substrate. And the faster degradation of few-layer GaSe on SiO$_2$ may be caused by the O$_2$ in the substrate and slow thermal dissipation.

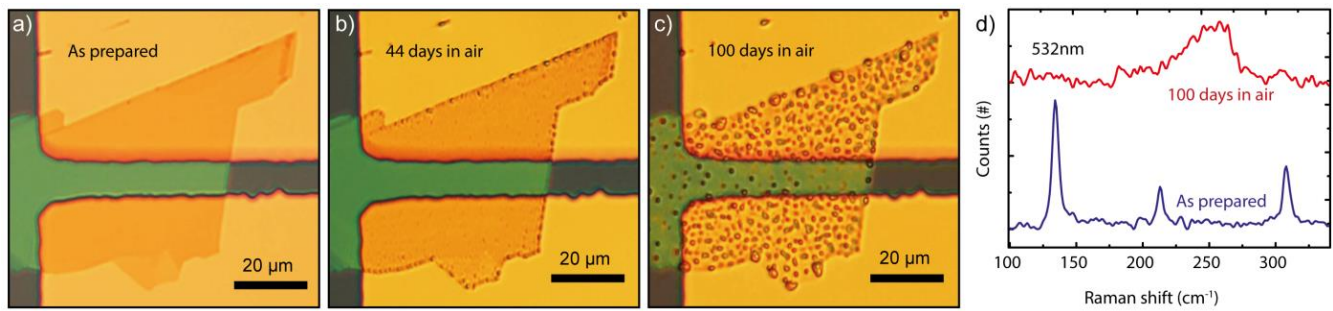

**Figure S5**: **Evolution of a GaSe device in air.** Optical images of the GaSe device kept in air discussed in main text just after fabrication (a) and after 44 (b) and 100 (c) days in air. d) Corresponding Raman spectra of the GaSe device in (a) and (c).

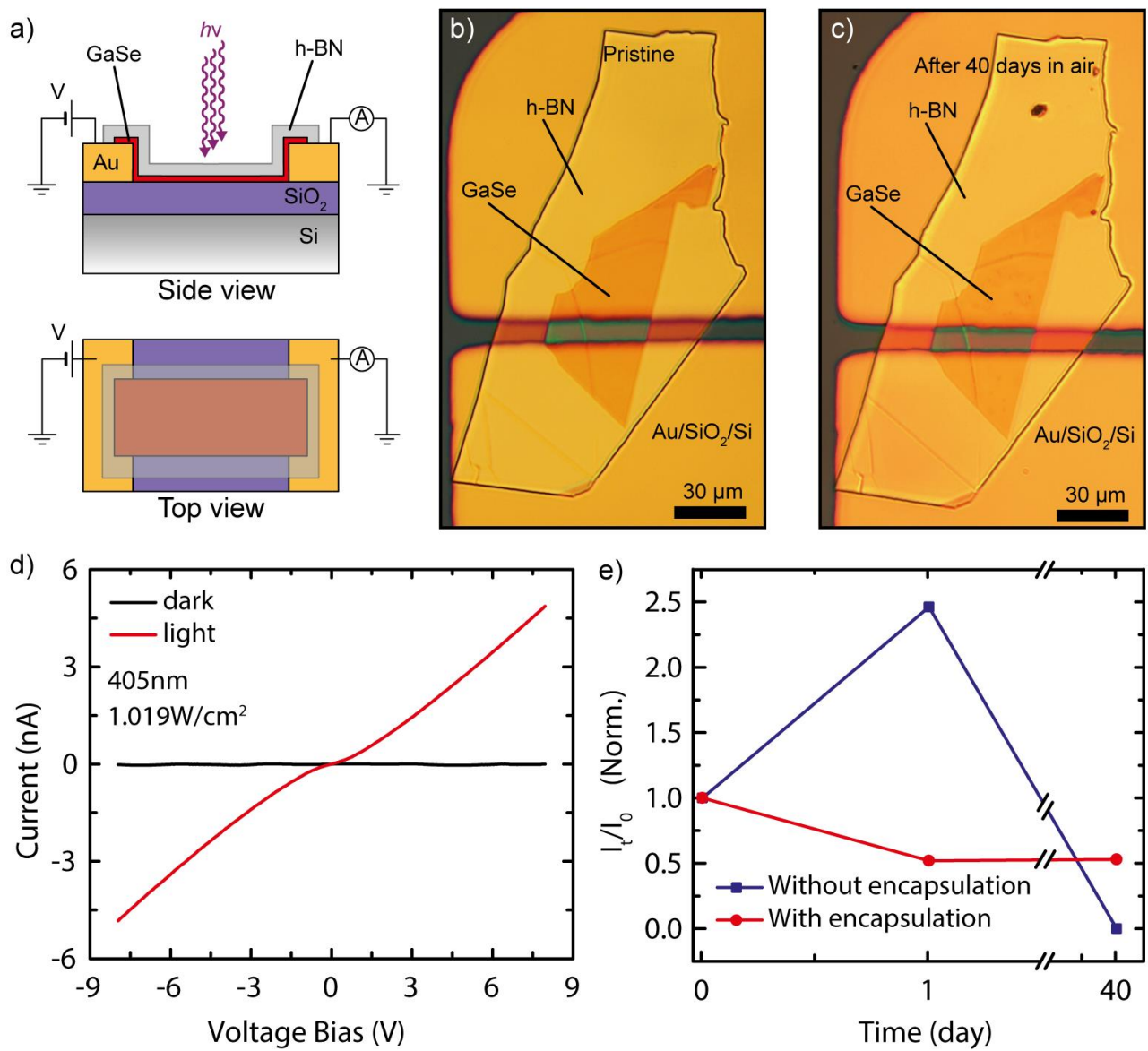

**Figure S6: h-BN encapsulated device.** a) Schematic of a h-BN encapsulated GaSe photodetector. b-c) Optical micrographs of a h-BN encapsulated GaSe photodetector recorded just after fabrication (b) and after 40 days in air (c). d) *I-V* characteristic of the GaSe photodetector encapsulated with h-BN after 40 days in air. c) The long-term photocurrent evolution of GaSe photodetector with and without h-BN encapsulation in air.